\documentclass[aps,prx,onecolumn,superscriptaddress]{revtex4-2}
\usepackage{graphicx}
\usepackage{xcolor}
\usepackage[normalem]{ulem} 
\usepackage{array}
\usepackage{amsmath}

\newcommand{\ket}[1]{|#1\rangle}

\begin{document}


\title{Coherent Control of an Optical Quantum Dot Using Phonons and Photons}


\author{Ryan A DeCrescent}
\email{ryan.decrescent@nist.gov}
\affiliation{National Institute of Standards and Technology, Boulder, Colorado 80305, USA}
\author{Zixuan Wang}
\affiliation{National Institute of Standards and Technology, Boulder, Colorado 80305, USA}
\affiliation{Department of Physics, University of Colorado, Boulder, CO 80309, USA}
\author{Joseph T Bush}
\affiliation{National Institute of Standards and Technology, Boulder, Colorado 80305, USA}
\affiliation{Department of Physics, University of Colorado, Boulder, CO 80309, USA}
\author{Poolad Imany}
\affiliation{National Institute of Standards and Technology, Boulder, Colorado 80305, USA}
\affiliation{Department of Physics, University of Colorado, Boulder, CO 80309, USA}
\affiliation{Icarus Quantum Inc., Boulder, CO 80302}
\author{Alex Kwiatkowski}
\affiliation{National Institute of Standards and Technology, Boulder, Colorado 80305, USA}
\author{Dileep V Reddy}
\affiliation{National Institute of Standards and Technology, Boulder, Colorado 80305, USA}
\author{Sae Woo Nam}
\affiliation{National Institute of Standards and Technology, Boulder, Colorado 80305, USA}
\author{Richard P Mirin}
\affiliation{National Institute of Standards and Technology, Boulder, Colorado 80305, USA}
\author{Kevin L Silverman}
\email{kevin.silverman@nist.gov}
\affiliation{National Institute of Standards and Technology, Boulder, Colorado 80305, USA}

\date{\today}

\begin{abstract}
Genuine quantum-mechanical effects are readily observable in modern optomechanical systems comprising \emph{bosonic} (``classical") optical resonators. Here we describe unique features and advantages of optical \emph{two-level systems}, or qubits, for optomechanics. The qubit state can be coherently controlled using \emph{both} phonons and resonant or detuned photons. We experimentally demonstrate this using charge-controlled InAs quantum dots (QDs) in surface-acoustic-wave resonators. Time-correlated single-photon counting measurements reveal the control of QD population dynamics using engineered optical pulses and mechanical motion. As a first example, at moderate acoustic drive strengths, we demonstrate the potential of this technique to maximize fidelity in quantum microwave-to-optical transduction. Specifically, we tailor the scheme so that mechanically assisted photon scattering is enhanced over the direct detuned photon scattering from the QD. Spectral analysis reveals distinct scattering channels related to Rayleigh scattering and luminescence in our pulsed excitation measurements which lead to time-dependent scattering spectra. Quantum-mechanical calculations show good agreement with our experimental results, together providing a comprehensive description of excitation, scattering and emission in a coupled QD-phonon optomechanical system.
\end{abstract}


\maketitle

Optomechanics underlies major technological advancements ranging from single particle and atom trapping \cite{ashkin_optical_1997} to gravitational-wave detection \cite{aspelmeyer_cavity_2014, blair_optomechanics_2020}. As detection sensitivities and fabrication capabilities have improved, cavity optomechanics has been extended to explore and exploit more subtle quantum-mechanical effects \cite{aspelmeyer_cavity_2014, barzanjeh_optomechanics_2022}, enabling laser-cooling of mesoscopic objects \cite{schliesser_resolved-sideband_2008, chan_laser_2011, brubaker_optomechanical_2022, saarinen_laser_2023}, new sources of squeezed light \cite{purdy_strong_2013}, transduction between microwaves and optical light \cite{andrews_bidirectional_2014, balram_coherent_2016, mirhosseini_superconducting_2020, forsch_microwave--optics_2020, brubaker_optomechanical_2022, weaver_integrated_2024}, and entanglement between optical and mechanical \cite{riedinger_non-classical_2016, rakhubovsky_detecting_2020} or optical and microwave modes \cite{jiang_optically_2023, meesala_quantum_2023}. Notably, all such previous demonstrations have exploited \emph{bosonic} (``classical") optical resonators, for example, whispering-gallery-mode resonators \cite{schliesser_resolved-sideband_2008}, Fabry-Perot resonators \cite{andrews_bidirectional_2014, saarinen_laser_2023}, and photonic-crystal nanobeams \cite{chan_laser_2011, balram_coherent_2016, mirhosseini_superconducting_2020, forsch_microwave--optics_2020, meesala_quantum_2023, weaver_integrated_2024}; their purpose is essentially to confine the optical field to the region where the optomechanical interaction occurs, or to enforce frequency-dependent optical admittances that depend sensitively on mechanical motion. Genuine quantum control of mechanical motion, however, requires an external \emph{two-level system} (TLS) for state preparation and readout. This has in fact been realized in \emph{microwave} electromechanics by coupling superconducting qubits to mechanical modes, sparking the field of quantum acoustodynamics \cite{satzinger_quantum_2018,chu_creation_2018}. Analogous interactions have not yet been established between mechanical modes and \emph{optical} TLSs, although this would be beneficial in applications such as quantum microwave-optical transduction, single-phonon generation \cite{sollner_deterministic_2016} or single-photon generation \cite{bracht_swing-up_2021}. Optical TLSs inherently limit optical interactions to single-photon levels and enable coherent control protocols that are not possible in bosonic optical resonators. Optical TLSs and related \emph{three}-level systems can resonantly couple with mechanical modes, offering potential routes to efficient optomechanical cooling and phonon ``lasing" \cite{kepesidis_phonon_2013}.

Semiconductor quantum dots (QDs) are excellent realizations of optical TLSs \cite{ester_ramsey_2006, press_complete_2008}, exhibiting nearly gigahertz emission rates, near-unity quantum yields, near-transform-limited linewidths, high single-photon purity and high single-photon indistinguishability \cite{senellart_high-performance_2017}. They have been used to generate nonclassical states of light \cite{trivedi_generation_2020}, spin-photon entanglement \cite{luo_spinphoton_2019, appel_entangling_2022, coste_high-rate_2023}, and photon-photon entanglement \cite{chen_highly-efficient_2018, liu_solid-state_2019}. Demonstrations of unconventional coherent excitation schemes using two ultrafast red-detuned laser pulses highlight the remarkable degree of control possible with QDs \cite{bracht_swing-up_2021, karli_super_2022}. Recent efforts involve integrating QDs into larger multifunctional hybrid quantum systems \cite{lodahl_interfacing_2015, tsuchimoto_large-bandwidth_2022, buhler_-chip_2022, larocque_tunable_2023}. QDs and other solid-state quantum emitters, e.g., diamond color centers, have more recently been explored as light-matter intermediaries in quantum optomechanical systems \cite{kepesidis_phonon_2013, schuetz_universal_2015, weis_interfacing_2018, maity_coherent_2020, kettler_inducing_2020, lukin_spectrally_2020, wigger_resonance-fluorescence_2021, imany_quantum_2022, decrescent_large_2022, hahn_photon_2022, wang_gated_2023, spinnler_single-photon_2023, patel_surface_2024}. Oscillations between excited orbital states in diamond color centers have been observed using multi-phonon processes and introduced as a method for dynamically mitigating decoherence \cite{mccullian_coherent_2024}. To date, we know of no demonstrations of coherent optical and mechanical control of a coupled QD-phonon system. 

Here, QD excitation dynamics are controlled using \emph{both} phonons and photons. Mechanical motion has large effects on Bloch-sphere trajectories up to approximately 1 ns after the onset of optical driving, even after substantial radiative relaxation. As a first example, we establish a protocol directly applicable to optically heralded entanglement of remote acoustic resonators. In such a protocol, detection of an optomechanically scattered photon heralds the successful preparation of the mechanical state, but zero-phonon scattering events are ``false alarms" that degrade state preparation fidelity. We show that optical pulse shapes can be optimized such that phonon-assisted scattering is substantially enhanced compared to background scattering from direct detuned excitation, mitigating the deleterious effect of such background photons. There is no analogue in optomechanical systems comprising classical optical resonators. Finally, we analyze the frequency spectrum of photons from a QD prepared using either direct or phonon-assisted excitation. These spectra reveal distinct Rayleigh-like scattering channels and luminescence channels which in the pulsed-excitation case lead to time-dependent spectra and verify successful optomechanical excitation of the QD. These results provide a comprehensive picture of excitation, scattering, and emission processes in a coupled phonon-QD optomechanical system. 

\section{Understanding phonon-assisted TLS dynamics}
Current realizations of hybrid QD-mechanical systems are operated at timescales much longer than relevant dephasing times, i.e. via continuous-wave (CW) pumping \cite{lukin_spectrally_2020, wigger_resonance-fluorescence_2021, tsuchimoto_large-bandwidth_2022, imany_quantum_2022, spinnler_single-photon_2023}, or with nonresonant optical pumping \cite{couto_photon_2009, weis_interfacing_2018, decrescent_large_2022}, and therefore these coherent properties are obscured. In a few cases, unique TLS signatures have been observed through second-order photon correlation measurements \cite{lukin_spectrally_2020, spinnler_single-photon_2023}. Dynamics in phonon-assisted scattering have been observed, but the optical TLS coherence was not exploited or explored \cite{wigger_resonance-fluorescence_2021}. In the CW case, the QD merely plays a role analogous to the optical resonators in the above-mentioned bosonic systems. The two-level nature of the QD could even be considered a disadvantage in these cases, as saturation can occur under the strong optical pumping required for significant optomechanical coupling rates. This is remedied by using short optical pulses and well-known coherent control methods, as we show here.

Our system comprises a single optical QD with transition frequency $\omega_\text{QD}$ subject to an optical drive of frequency $\omega_\text{pump}$=$\omega_\text{QD}$+$\Delta$ with detuning $\Delta$. It is coupled to phonons in a single-mode SAW cavity at frequency $\omega_\text{SAW}$ (Fig. 1a). The SAW strain field parametrically modulates the QD frequency via deformation potential coupling \cite{metcalfe_resolved_2010, schuetz_universal_2015,  golter_coupling_2016, aref_quantum_2016}. Including this coupling, the QD-SAW system can be described by a ladder-like energy diagram (Fig. 1b). Diagonal transitions that add (remove) phonons to (from) the acoustic resonator are addressed with an optical drive at detuning $+\omega_\text{SAW}$ ($-\omega_\text{SAW}$) with a renormalized interaction rate of $\Gamma$ \cite{aspelmeyer_cavity_2014, golter_coupling_2016}. Even with this detuning, however, there is a finite probability of exciting the QD ``directly", i.e., via vertical transitions involving no phonons. This process is inherent to the TLS and is described by the generalized Rabi rate $\Omega$=$\sqrt{\Omega_0^2+\Delta^2}$, where $\Omega_0$ is the resonant Rabi rate between the QD and pump field. This undesired excitation channel can be removed by controlling the system in the time domain. Fig. 1c presents simplified calculations illustrating the protocol. The zero-phonon (``direct") excitation channel leads to occupancy oscillations of the $\ket{e,n}$ state at frequency $\Omega$ (Fig. 1c, black curve). Excitation via the phonon-assisted transition, which is in fact resonant with the pump, evolves differently (red curve). There are thus discrete pulse durations where the direct excitation channel is effectively eliminated, leaving only the phonon-assisted excitation. This phonon-assisted excitation process can alternatively be visualized on the bare QD's Bloch sphere (Fig. 1d). Without phonon coupling, the QD’s state traces closed small circles near the south pole (ground state) of the Bloch sphere (black). The acoustic field perturbs this trajectory, generally increasing the QD’s occupancy while the optical drive remains active. The QD occupancy approaches unity given enough time and assuming sufficiently slow dephasing \cite{kuniej_hybrid_2024}. Further theoretical details are provided in the Supplementary Information.

\begin{figure}
      \includegraphics[width=0.5\textwidth]{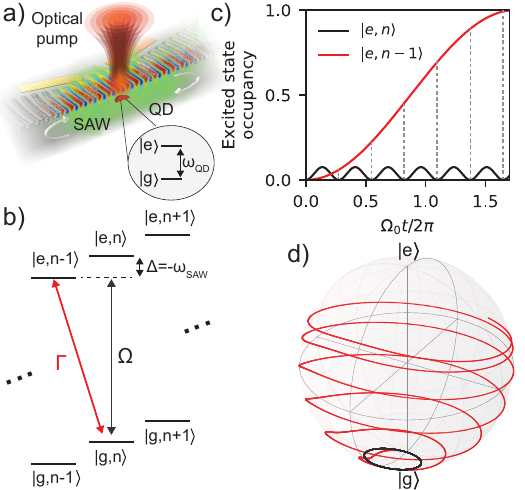}
	\caption{
        (\textbf{a}) Illustration of the system. A QD is coupled to surface acoustic wave (SAWs) cavity phonons and is driven by an optical pump.
        (\textbf{b}) Schematic energy-level ladder diagram of the coupled QD and phonon resonator (frequency $\omega_\text{SAW}$). A pump laser with frequency detuning $\Delta$=$-\omega_\text{SAW}$ ($+\omega_\text{SAW}$) can address optical-frequency transitions by removing (adding) phonons to (from) the resonator. The coupling rate associated with these diagonal transitions is $\Gamma$=$g_0 \Omega_0 \sqrt{n} / \omega_\text{SAW}$, where $g_0$ is the single-phonon optomechanical coupling rate, $n$ is the cavity phonon number, and $\Omega_0$ ($\Omega$=$\sqrt{\Omega_0^2 + \Delta^2}$) is the resonant (generalized) Rabi rate between the laser field and the QD.
        (\textbf{c}) Calculated excited-state occupancies of the states $\ket{e,n}$ (black) and $\ket{e,n-1}$ (red) from a square pulse initiated at time $t$=0 with the following system parameters: $\Delta$=$-\omega_\text{SAW}$=$-$$2\pi\times$3.5 GHz, $g_0\sqrt n$=$2\pi\times$1 GHz, $\Omega_0$=$2\pi\times$1 GHz. Vertical gray dashed lines indicate pulse durations where high-fidelity optomechanical state preparation is possible.
        (\textbf{d}) Bare QD Bloch-sphere trajectories for direct (black) and phonon-assisted (red) excitation process with $\Delta$=$-\omega_\text{SAW}$=$-$$2\pi\times$3.5 GHz, $g_0\sqrt n$=$2\pi\times$2 GHz, $\Omega_0$=$2\pi\times$1 GHz. Calculations in panels b) and c) assume no dephasing.
	}
\label{fig:1}
\end{figure}

\section{Characterizing basic device performance}
A fabricated QD-SAW optomechanical device is shown in Fig. 2a. (Complete fabrication details are described in Supplementary Information.) SAWs are confined between two acoustic mirrors forming an acoustic Fabry-Perot cavity at frequency $\omega_\text{SAW}/2\pi$$\approx$3.6 GHz with a quality factor $Q$$\approx$20,000 and finesse $\approx$130. The cavity both enhances the QD-SAW interaction compared to that on a bare surface \cite{imany_quantum_2022} and allows us to parameterize the QD-SAW interaction by the single-phonon coupling rate $g_0$ and equilibrium phonon number $\bar{n}_\text{SAW}$. The SAW cavity mode is coherently populated with $\bar{n}_\text{SAW}$ phonons by driving an interdigital transducer (IDT) with an external microwave source. Here, we use large SAW cavities with small $g_0/2\pi$$\approx$10 kHz. Nonetheless, $g$=$\sqrt{\bar{n}_\text{SAW}}g_0$ reaches $g/2\pi$$\gtrsim$1 GHz with phonon occupancies of $\bar{n}_\text{SAW}$$\sim$$10^9$ which we readily achieve with modest external microwave driving powers of approximately $-45$ dBm. Previous work has reported single-phonon optomechanical coupling rates $g_0$ as large as 1-3 MHz \cite{decrescent_large_2022, spinnler_single-photon_2023} wherein the effects demonstrated here should be readily observable with $\sim$$10^3$ phonons. 

\begin{figure}
      \includegraphics[width=0.5\textwidth]{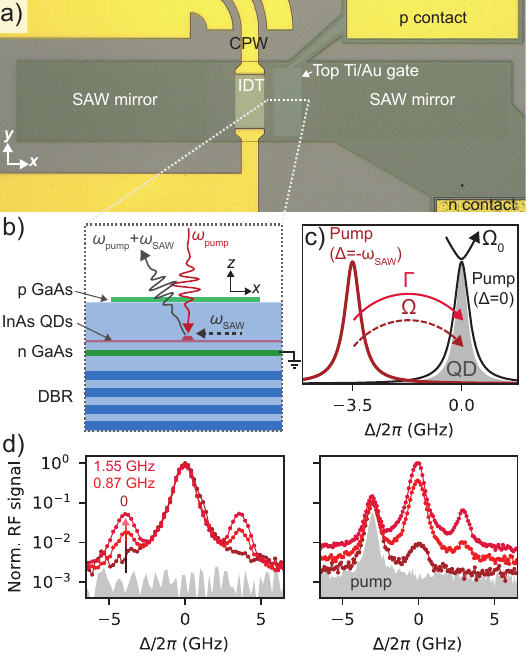}
	\caption{	
	(\textbf{a}) Optical microscope image of the device. IDT: interdigital transducer. CPW: coplanar waveguide. 
	(\textbf{b}) Schematic cross-section (not to scale) of the region under the p-type top gate (indicated by a dotted white line in panel a). A single InAs QD mediates an optomechanical interaction between SAWs ($\omega_\text{SAW}$) and the optical pump ($\omega_\text{pump}$). DBR: distributed Bragg reflector.
	(\textbf{c}) Frequency-domain illustration of the system. Three different scattering processes are illustrated with rates $\Omega_0$ (resonant direct excitation), $\Omega$ (detuned direct excitation) and $\Gamma$ (detuned phonon-assisted excitation), described in the main text. 
	(\textbf{d}) Two types of resonance fluorescence (`RF') measurements from a single QD. Left panel: Total scattered photon counts while slowly varying the CW pump detuning $\Delta$ without (brown) and with large steady-state phonon occupancies ($g/2\pi$=0.87 GHz, dark red; $g/2\pi$=1.55 GHz, lighter red). Right panel: Frequency spectrum of scattered photons under red-detuned ($\Delta$=$-\omega_\text{SAW}$=$-2\pi\times$3.58 GHz) CW optical pumping. The $g$ values and corresponding colors are the same in both panels. In each panel, spectra have been normalized to the maximum counts in the dataset. Gray regions: measured pump background.
	}
\label{fig:2}
\end{figure}

For good QD performance, QDs are electrostatically stabilized and controlled by a vertical p-i-n diode (Fig. 2b) \cite{reigue_resonance_2019, wang_gated_2023}. This enables tuning of the QD’s charge state and wavelength with applied gate bias voltage (Fig. S2) \cite{hogele_voltage-controlled_2004, lobl_narrow_2017} and reduces QD absorption linewidths to $\gamma_\text{QD}/2\pi$$<$350 MHz in best cases \cite{matthiesen_subnatural_2012, kuhlmann_transform-limited_2015, lobl_narrow_2017}. We interrogate a singly charged QD at 956.368 nm with an approximately 950 MHz linewidth (full width at half maximum) measured over a timescale of several seconds. This broadening beyond the radiative lifetime limit is likely due to slow (sub-MHz) spectral wandering \cite{kuhlmann_charge_2013, kuhlmann_transform-limited_2015} rather than additional dephasing at the timescale of the QD's lifetime (Supplementary Information). Nonetheless, our system satisfies the condition $\omega_\text{SAW}$$>$$\gamma_\text{QD}$, known as the resolved-sideband regime. Fig. 2c illustrates the spectral components and the three scattering processes (described in the previous section) of primary concern in our experiments. Operation in the resolved-sideband regime is verified with CW optical spectroscopies by measuring the absorption (Fig. 2d; left panel) and scattering spectra (Fig. 2d; right panel) of the single QD with and without SAW modulation. Well-defined spectral sidebands at multiples of $\omega_\text{SAW}$ correspond to single-phonon-assisted optical scattering from the QD \cite{metcalfe_resolved_2010, weis_optomechanical_2021, imany_quantum_2022}. To measure system dynamics, we excite the QD with frequency-tuned and time-tailored optical pulses (Methods). A CW pump with variable detuning $\Delta$ originates from a narrow-linewidth ($\sim$10 kHz) laser. Pulses with durations of approximately 100 ps to $>$1 ns are generated by passing the CW pump through two electro-optic modulators and a tunable spectral filter (Methods). The time distribution of scattered photons is resolved using time-correlated single-photon counting (TCSPC) techniques.

\section{Experimentally resolving phonon-assisted excitation dynamics}
Using near-resonant pulses with duration similar to the QD lifetime, we directly observe the QD’s average population dynamics as it emits photons at random times during each pulse. These time-domain traces of the QD's dynamics are each composed of several million photon detection events (Methods). Series of such measurements under red-detuned ($\Delta$=$-\omega_\text{SAW}$) pumping with $g$ varying between approximately 0 (thermal phonon occupancy) and 1.55 GHz are shown in Figs. 3a-b. Measured count rates are converted to estimated QD occupancies using a calibration procedure (Fig. S3). For the remainder of this article, we will refer to the ratio of occupancies (minus one) as the ``enhancement", $c_g(t)$=$s_g(t)/s_{g=0}(t) - 1$ where $s_g(t)$ is the photon signal at time $t$ with coupling rate $g$. This enhancement measures how well the presence or absence of phonons can be distinguished with the optical signal. If operated in the quantum regime, it is directly related to fidelity.

\begin{figure*}
      \includegraphics[width=0.95\textwidth]{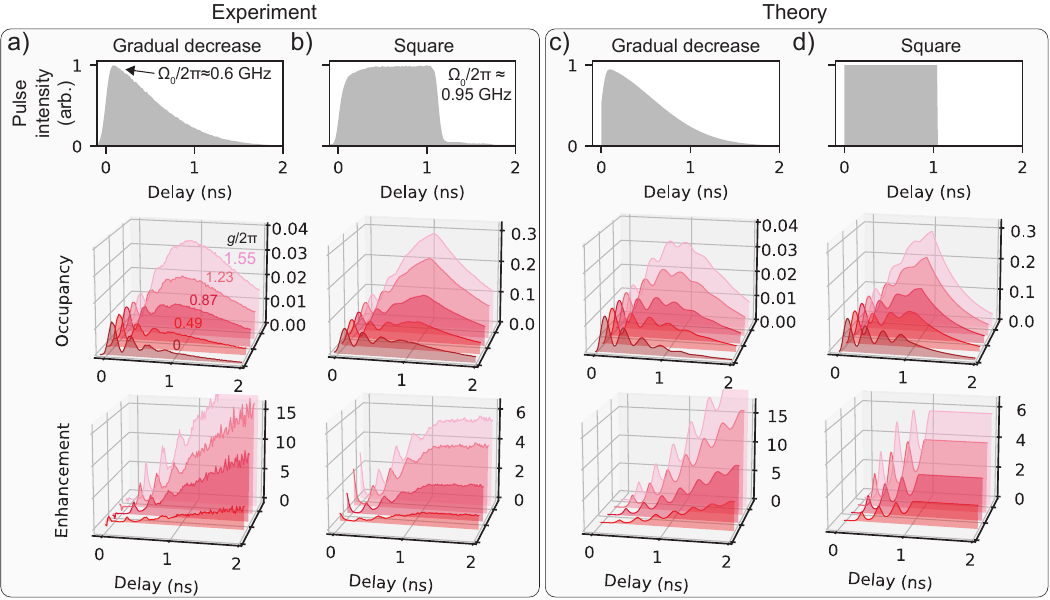}
	\caption{	
	(\textbf{a,b}) Experimental QD dynamics under red-detuned ($\Delta$=$-\omega_\text{SAW}$) pumping with two different optical pulse shapes: 
	a) a pulse with a gradually decreasing intensity; 
	b) a nearly square pulse with an abruptly decreasing intensity. Top row: measured pulse intensities for the two cases. Approximate peak resonant Rabi frequencies are specified in each panel. Center row: measured time-resolved QD occupancy. Trajectories for a range of QD-SAW coupling rates $g$ (different colors) are shown ($g/2\pi$ specified next to each curve in units of GHz). Bottom row: ``enhancements", $c_g(t)$, for each of the $g$$>$0 curves compared to the $g$=0 curve from the occupancy data shown in the center row. 
	(\textbf{c,d}) Numerical calculations corresponding to the two experimental scenarios presented in panels a,b. Calculations assume radiative relaxation at $2\pi\times$320 MHz and small additional dephasing ($2\pi\times$60 MHz) to match the experiment. Fig. S5 shows the same data with a two-dimensional format.
	}
\label{fig:3}
\end{figure*}

We first discuss measured dynamics using a pulse with a gradually decreasing intensity (Fig. 3a). Experimentally measured pulse shapes are shown in the top panels of Fig. 3a. The QD occupancy (Fig. 3a, center panel) with no added phonons ($g/2\pi$=0, brown curve) shows several Rabi oscillations at frequency $\Omega$, which is dominated by the detuning. Obvious changes in the dynamics are resolved when the SAW cavity is occupied such that $g/2\pi$=$490$ MHz. These differences become more pronounced as $g$ increases, where the QD occupancy is estimated to reach 0.035 for $g/2\pi$=$1.55$ GHz at 900 ps. The data reveal that there are several pulse durations where phonon-assisted excitation (red curves) dominates substantially over the bare QD excitation (brown curve) --- namely around 300 ps, 550 ps, and 800 ps --- largely governed by times when the bare QD occupancy swings down toward zero. These measured enhancements $c_g(t)$ (bottom panel) reach approximately 15 at 1.1 ns --- and continue to increase smoothly as the pulse slowly wanes --- but vary widely depending on the SAW cavity occupancy and pulse duration. To illustrate the importance of the pulse shape on the measured enhancements, we perform a similar set of measurements using an (approximately) square optical pulse (Fig. 3b). Although obvious changes in the dynamics are again observed when $g/2\pi$$>$0, the measured enhancements are considerably smaller than in the ``decreasing pulse" case.

The benefit of a \emph{gradually} decreasing pulse intensity --- as compared to a square pulse, for example --- becomes clear upon inspection of the bare QD dynamics. Numerical calculations of QD dynamics are presented in Figs. 3c-d, corresponding to the two experimental cases shown in Figs. 3a-b (Methods). For a pulse with constant intensity (Fig. 3d), the bare QD occupancy oscillates around and approaches a constant central value (here, 0.06) as dephasing proceeds (brown curve). Consequently, successive minima in the bare QD’s occupancy grow increasingly \emph{large} as the pulse continues. This behavior is well-resolved in our experiments (Fig. 3b). Enhancements thus remain small throughout the pulse duration. If the pulse intensity is instead gradually decreased soon after the onset, minima in the bare QD’s occupancy do \emph{not} increase over the pulse duration (Fig. 3c; brown curve). Enhancements thus \emph{increase} over the course of the QD’s lifetime or longer as the phonons have more time to interact with the optically driven QD.

There is a competition between the rate of decrease of the pump intensity and the QD dephasing. An ideal pulse would bring the QD occupancy down to zero before substantial dephasing. This can be accomplished by improving the QD coherence time, or by decreasing the pump intensity quicker than that demonstrated in Fig. 3a. Calculations presented in Fig. 3c,d assume radiative relaxation and small additional dephasing to match the experimental dynamics, although this additional dephasing likely imitates spectral wandering in this particular case. Reduced dephasing and no spectral wandering in an otherwise identical system substantially improves the enhancements (Fig. S6). With no dephasing nor spectral wandering, enhancements approach maximum values of approximately 1000 for $g/2\pi$$\gtrsim$1 GHz around 1.3 ns delay as the bare QD occupancy grows increasingly close to zero over the duration of the gradually decreasing pulse. 

In our measurements, successive detection events sample different mechanical phases $\phi$, and so the occupancies shown in Fig. 3a,b represent the dynamics averaged over all such phases. Calculations shown in Figs. 3c,d are thus averaged over this phase. While $\phi$ does affect the specific trajectory, it has only a weak effect on the shape and magnitude of $c_g(t)$ which is our primary concern here (Supplementary Information). This scenario would in fact be relevant for, e.g., single-mode quantum microwave-to-optical transduction where the relevant phonon Fock states do not carry a well-defined phase. For applications where the specific trajectory should be well controlled, for example in the above-mentioned ``swing-up" scheme, one can synchronize the mechanical drive phase with the onset of the optical pump \cite{kuniej_hybrid_2024}.

\section{Distinguishing Scattering Channels During Phonon-Assisted Pulsed Excitation}

The spectrum of photons scattered from the coupled QD-SAW system carries information about the state of both the SAW resonator and the QD. In our time-resolved measurements presented above, we inferred the QD occupancy from the total number of photons scattered or emitted over \emph{all} frequencies. Those measurements thus generally do not distinguish between various scattering mechanisms including, e.g., Rayleigh scattering and QD spontaneous emission. To address this, we prepare our system using two distinct red-detuned ($\Delta$=$-\omega_\text{SAW}$) square pulses (Fig. 4a). This state preparation is visualized through the QD dynamics both without SAW driving (gray curves) and with moderate SAW driving (orange curves). The scattered photons are then frequency resolved with filters of varying transmission bandwidths, $\delta f$ (Fig. 4b). Pulse durations are chosen to provide a maximum contrast between the two spectra; the shorter pulse (duration $\tau_\text{pulse}$$\approx$370 ps) ends at a maximum of the coherent zero-phonon Rabi cycle (Fig. 4a; left panel), and the longer pulse ($\tau_\text{pulse}$$\approx$500 ps) ends at the next minimum (right panel). 

Spectra measured under the two different pulse scenarios with and without SAW driving are displayed in Fig. 4c. The first notable feature is the weak peak at the laser drive frequency ($\Delta$=$-\omega_\text{SAW}$). We attribute this to Rayleigh scattering, identified by the spectral widths which are characteristic of the optical pulses for each pulse type, and by the little variation from case to case. The resonance at $\Delta$=0 is more interesting. Comparing spectra with and without SAW driving, we identify components related to both the detuned zero-phonon excitation of the QD and the phonon-assisted excitation, depending on the preparation protocol. With the short pulse --- which can be considered a ``poor" choice for isolating optomechanical excitation --- both spectra are almost completely dominated by the zero-phonon excitation process. With the longer pulse (right two panels) most of the photons are from the phonon-assisted excitation process (orange). These results are in fact the experimental realization of the process illustrated in Fig. 1c.

\begin{figure}
      \includegraphics[width=0.5\textwidth]{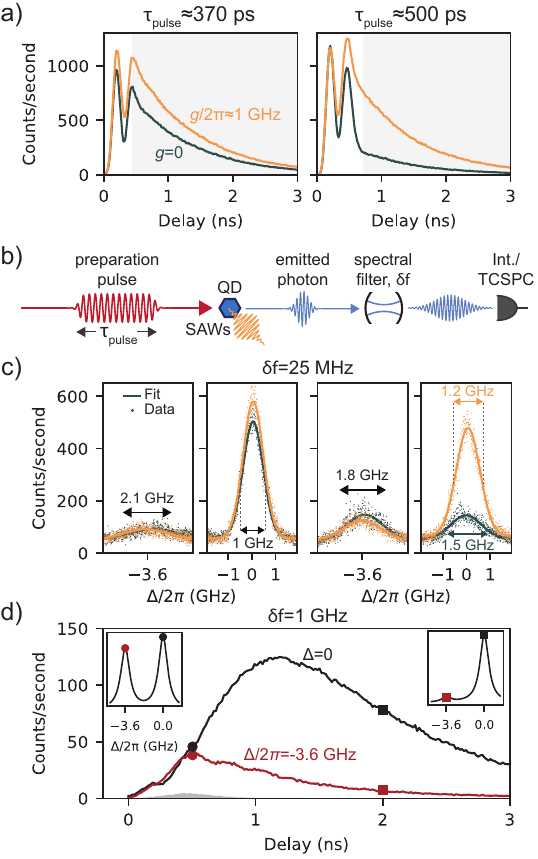}
	\caption{	
	(\textbf{a}) Experimental time-resolved photon signal with red-detuned ($\Delta$=$-\omega_\text{SAW}$=$-2\pi\times$3.58 GHz) optical pulses of lengths $\tau_\text{pulse}$$\approx$370 ps (left panel) and $\tau_\text{pulse}$$\approx$500 ps (right panel). Gray curves: no SAW drive ($g$=0). Orange curves: SAW drive ($g$$\approx$1 GHz).
	(\textbf{b}) Schematic experimental setup for measuring the spectra shown in panels b) and c). After pulsed excitation with and without SAW driving, collected photons are passed through a tunable spectral filter with transmission bandwidth $\delta f$ and measured either by integrating over all photon arrival times (`Int.') or by recording arrival times with TCSPC.
	(\textbf{c}) Spectra measured with $\delta f$=25 MHz. Left two panels correspond to $\tau_\text{pulse}$=370 ps. Right two panels correspond to $\tau_\text{pulse}$=500 ps. Panels are split to highlight two important spectral features around $\Delta$=$-\omega_\text{SAW}$ and $\Delta$=0. The spectral width of each subpanel is identical (3.8 GHz). Data (dots) have been fit to Gaussian functions (solid curves). Text specifies the fit full-width at half maximum.
	(\textbf{d}) Time-resolved photon signal (same setup as in panel a) measured with $\delta f$=1 GHz around $\Delta/2\pi$=$-$3.58 GHz (red curve) and $\Delta$=0 (black curve). The gray region is the measured pump background. Insets show reconstructed spectra at the 0.5 ns (circle markers) and 2.0 ns (square markers). 
	}
\label{fig:4}
\end{figure}

More information is obtained by measuring the time distribution of collected photons after passing them through a $\delta f$=1 GHz spectral filter (Fig. 4d). This allows the pump and QD frequencies to be isolated while retaining temporal resolution on the order of $1/\delta f$$\approx$1 ns; time traces become ``smeared" over about 1 ns. The decay of the $\Delta$=$-\omega_\text{SAW}$ component (red curve) follows the abrupt termination of the excitation pulse, while the $\Delta$=0 component (black curve) decays much more slowly, at approximately the QD lifetime. This further confirms Rayleigh scattering as the origin of most of the the red-detuned photons; QD preparation and subsequent luminescence accounts for the emission at $\Delta$=0. These different temporal dynamics are consistent with the spectral widths of the peaks measured in Fig. 4c. From this data, we make artificial reconstructions of the scattering spectrum at 0.5 ns (Fig. 4d; left inset, circle markers) and 2 ns (right inset; square markers). The spectrum varies in time during and after the pulsed excitation, largely resembling the QD at long times after the pump terminates, confirming the successful preparation of the QD state. 

\section{Discussion}
We experimentally demonstrates and theoretically described phonon-assisted excitation dynamics in a semiconductor QD using detuned optical pulses and coherent phonon driving. Specifically, time-correlated single-photon counting measurements reveal excitation dynamics that depend on the presence or absence of microwave-frequency phonons. We showed that the QD can be prepared in an excited state preferentially by phonon-driven processes with strategic optical pulse shaping. That is, the specific dynamics are sensitive to the presence of phonons and details of the optical pulse, including amplitude, temporal envelope, detuning, and phase. Complementary frequency-domain measurements reveal time-dependent scattering channels related to Rayleigh scattering and luminescence, and verify the successful preparation of the QD excited state. These behaviors may be exploited for applications such as microwave-to-optical transduction where photon scattering unrelated to phonon-assisted processes is highly undesirable. In this context, we demonstrated that pulses with a gradually decreasing intensity are beneficial. More complicated pulse shapes involving frequency, amplitude and phase shaping may be developed to further improve this performance. 

More generally, this work shows that with the QD, new parameters are available for interacting with the mechanical system at nanosecond timescales. With this time-domain control, there is no fundamental reason for matching the detuning to the mechanical mode frequency as is needed with bosonic resonators \cite{aspelmeyer_cavity_2014}. Using additional detuned pumps may offer another level of control, similar to previously demonstrated ``swing-up" schemes \cite{bracht_swing-up_2021, karli_super_2022, kuniej_hybrid_2024}. Since the QD’s lifetime sets the timescale of state preparation and readout, average optical pump powers used for these protocols are at the nanowatt scale and this optical power is not stored in a resonator, reducing absorption-induced heating which is a major concern in optomechanical crystals \cite{meenehan_silicon_2014, forsch_microwave--optics_2020, jiang_optically_2023}. 

Comparing our experimental results to numerical calculations suggests that our measured QD experiences slow spectral fluctuations that hinder the ability to repeatedly prepare the QD in a desired state over many seconds. Modern optimized charge-controlled QDs regularly exhibit near-lifetime-limited linewidths \cite{kuhlmann_transform-limited_2015, lobl_narrow_2017, tomm_bright_2021}, and we thus anticipate that performance comparable to our idealized numerical calculations is possible. Future efforts should involve optimizing QD performance in application-relevant mechanical resonators with micrometer-scale mode volumes and exploring the broad parameter space of optical pumping schemes for improving the enhancements described here.

\section{Acknowledgements}
We thank Emanuel Knill for informative discussions and intellectual support related to this work.

\section{Author Contributions}
R.A.D., A.K., and K.L.S. conceived and designed the experiments. R.A.D. performed the measurements and analyzed the data. All authors contributed materials and analysis tools. R.A.D. and K.L.S. wrote the paper with input from all other authors. 

\section{Methods}
\subsection{System Hamiltonian}
The Hamiltonian describing our system, in a frame rotating with the optical pump at frequency $\omega_\text{pump}$, assuming a semiclassical optical and mechanical drive is given by

\begin{equation}
\frac{\tilde{\hat{H}}(t)}\hbar = 
\frac12 \left[ -\Delta + g \cos(\omega_\text{SAW}t + \phi) \right]\hat{\sigma}_z 
+ \frac{\Omega_0(t)}2 \hat{\sigma}_x
\end{equation}

\noindent where $\Delta$, $\Omega_0$ and $g$ describe the pump-QD detuning, the resonant Rabi rate, and QD-SAW coupling, respectively. $\Omega_0$ is taken to be a real-valued function with a time dependence that describes the optical pulse envelope. We numerically calculate the dynamics of the system, described by Eqn. 1, using the Lindblad master equation with the open-source Python library QuTiP. Radiative decay is included through the Lindblad operator $\hat{\sigma}_- = (\hat{\sigma}_x+i\hat{\sigma}_y)/2$ with rate $\gamma_\text{QD}$. Additional pure dephasing is included through the Lindblad operator $\hat{\sigma}_z$ with rate $\gamma_z$. The rates $\gamma_\text{QD}$ and $\gamma_z$ are specified in the manuscript where relevant. Further details are provided in the Supplementary Information.

\subsection{Device Fabrication}
Full device fabrication details are provided in the Supplementary Information.

\subsection{Experimental Setup}
All of our QD optical measurements were performed using a 5 K optical cryostat with an internal cryogenic 0.81 NA objective. The resonant or detuned pump beam is shaped outside the cryostat using both single-mode optical fiber components and free-space optical components. The resonant pump beam was a tunable external-cavity diode laser with an approximate 10 kHz linewidth at 960 nm. Pump beams illuminate the QD through the objective. The QD signal was collected by the same objective and isolated from the pump predominantly by spectral filtering (for measurements involving a nonresonant or detuned pump) or polarization filtering (for measurements involving a resonant pump) with at least 10$^7$ pump rejection. Luminescence was collected into single-mode optical fiber and various frequency- and time-domain measurements were performed using procedures detailed below. The QD bias voltage was held constant for all measurements shown in the manuscript. The QD absorption/emission wavelength (and therefore the resonant pump wavelength) used in this work was 956.368 nm.

\subsection{Continuous-wave Resonance Fluorescence Spectra}
Continuous-wave (CW) resonance fluorescence (RF) spectra (e.g. Fig. 2d) were recorded by slowly sweeping the frequency of a tightly focused resonant pump beam over the QD’s exciton resonance while holding the QD bias constant. Luminescence count rates at each frequency were recorded using a superconducting nanowire single photon detector (SNSPD) mounted in a dilution refrigerator. A pump power of approximately 2 nW at the sample was used, corresponding to approximately 1/10 the measured saturation level of the QD. For manuscript Fig. 2d (left panel), such spectra were recorded without SAW driving and with SAW driving using an external coherent microwave source at approximately 3.587 GHz and external microwave driving powers between -50 dBm and -35 dBm. External microwave driving powers were converted to coupling rates $g$ by a series of calibration data where the modulation index $\chi$ was fit for each microwave driving power and $g$ was derived by the relationship $g$=$\omega_\text{SAW} \chi$. The calibration data are shown in the Supplementary Information.

For manuscript Fig. 2d (right panel), the pump was held at a constant frequency detuning of $\Delta$=$-\omega_\text{SAW}$. A pump power of approximately 5 nW was used. The frequency spectrum of scattered photons was measured by passing the collection port through a tunable Fabry-Perot etalon filter with a 600 MHz linewidth before being counted on an SNSPD. In manuscript Fig. 2d (right panel), such spectra were again measured without SAW driving and with SAW driving at 3.587 GHz and the same powers as in manuscript Fig. 2d (upper panel). Note that SAW driving frequencies varied by several MHz between measurements due to cavity frequency shifts over the duration of the measurements which spanned several days.

Further experimental details are provided in the Supplementary Information.

\subsection{Pulsed Optical Measurements and Time-domain QD Dynamics}
The time evolution of the QD dynamics was measured using standard time-correlated single-photon counting (TCSPC) techniques with a two-channel picosecond event timer and TCSPC module. For ``gradual decrease" pulses, optical pulses were generated by passing a CW detuned optical pump (same as described above) through two cascaded electro-optic modulators (EOMs). The EOMs were driven by amplified pulses from an approximately 100-ps square-wave electronic pulse generator (EPG). The EPG was externally triggered using an arbitrary waveform generator (AWG) at 34.085 MHz repetition rate. The rise time, fall time and width of the output pulses were measured to be approximately 15 ps, 15 ps, and 130 ps respectively. The on/off contrast was measured to be about $10^5$. The frequency bandwidth of the output pulses was measured to be 7 GHz. To decrease the bandwidth (and increase the pulse duration), the pulses were transmitted through a 600 MHz tunable Fabry-Perot etalon which was tuned for maximum transmission at the pulse’s center frequency. The resulting pulse shape is the ``gradual decrease" shape shown in Fig. 3. For square pulses, the EOMs were directly driven using a high-speed AWG (64 GS/sec) with approximately 30 ps rise time and pulses of arbitrary length. The pulse center frequency was detuned by $\Delta$ with respect to the QD’s resonance frequency (detuning specified in each measurement). 

The time distribution of the QD luminescence was measured by again detecting the luminescence with an SNSPD and time-correlating the counts on the TCSPC module with respect to a trigger signal from the AWG. We typically used 30-60 second total integration times, over which we collected typically several million total photons. This type of measurement was performed with a wide range of average optical input powers (between approximately 0.25 nW to 300 nW). Pump powers were converted to resonant Rabi rates by looking at the resulting time-domain oscillations with zero detuning. 

The time-resolved QD dynamics shown in the manuscript are presented after subtracting the experimentally measured pump contribution (i.e. unrejected pump background). Count rates were converted to approximate excited state occupancies using a series of power-dependent calibration data where observed Rabi oscillations were compared to numerical calculations. This subtraction process and scaling process are further illustrated in the Supplementary Information.

\subsection{Spectral measurements with pulsed excitation}
Spectral measurements shown in Fig. 4c,d were performed by collecting and passing the QD signal through two distinct frequency-tunable Fabry-Perot etalon filters. The pump configurations were identical in both cases, specified in the manuscript. For Fig. 4c, we used an etalon with a 25 MHz transmission bandwidth and present the data at each frequency (voltage) value. Fig. 4d, we used an etalon with a 1 GHz transmission bandwidth and present the resulting time-domain data for only two different transmission frequency windows corresponding to the (detuned) pump frequency and the QD frequency.

\bibliography{man}

\newpage
\renewcommand{\thefigure}{S\arabic{figure}}
\setcounter{figure}{0}
\renewcommand*{\thesection}{\arabic{section}}
\setcounter{section}{0}

\section*{Supplementary Information:\\Coherent Control of an Optical Quantum Dot Using Phonons and Photons}

\section{System Hamiltonian}
\subsection{General description}
Our system comprises a two-level system (TLS) with frequency $\omega_\text{QD}$ driven at optical frequency $\omega_\text{pump}$ with an amplitude described by the coupling (Rabi) rate $\Omega_0$. The TLS frequency is modulated at microwave frequency $\omega_\text{SAW}$ with a modulation depth $\chi$=$g/\omega_\text{QD}$. Here, we treat the optical drive and the mechanical drive semiclassically, which is appropriate for moderate SAW driving powers as used in our experiments. In this case, the phonon field amplitude is contained in the rate $g$, which can be related to the single-phonon coupling rate $g_0$ by $g=g_0\sqrt n$ when $n$ phonons occupy the resonator. The Hamiltonian $\hat{H}$ is:
\begin{equation}
\frac{\hat{H}(t)}\hbar = 
\frac12 \omega_\text{QD} \hat{\sigma}_z 
+ \frac12 g \cos(\omega_\text{SAW}t + \phi) \hat{\sigma}_z 
+ \Omega_0 \cos(\omega_\text{pump}t) \hat{\sigma}_x
\end{equation} 

We transform into a frame rotating about the TLS's $z$ axis at the pump frequency using the unitary operator $\hat{U}$=$e^{i\omega_\text{pump}\hat{\sigma}_z t / 2}$. The transformed Hamiltonian $\tilde{\hat{H}}$=$-i\hat{U}\dot{\hat{U}}^\dagger + \hat{U} \hat{H} \hat{U}^\dagger$, after applying the rotating-wave approximation for frequencies $\omega_\text{pump}$$\approx$$\omega_\text{QD}$ and identifying the pump-TLS detuning $\Delta$=$\omega_\text{pump} - \omega_\text{QD}$, is:
\begin{equation}
\frac{\tilde{\hat{H}}(t)}\hbar = 
\frac12 \left[ -\Delta + g \cos(\omega_\text{SAW}t + \phi) \right]\hat{\sigma}_z 
+ \frac{\Omega_0(t)}2 \hat{\sigma}_x
\end{equation}

We have allowed $\Omega_0$ to adopt a time dependence which describes the optical pulse envelope. Formally, Eqn. 2 assumes that the time variation of $\Omega_0(t)$ is slow enough that the resulting spectral distribution remains small compared to the detuning $\Delta$ so that the optical drive can be approximated as monochromatic. Experimentally, we enforce this using long pulses and/or spectral filters. We additionally assume that $\Omega_0(t)$ is a real-valued function. In general, pulse shaping will cause the pulse amplitude to become complex which may affect the specific TLS trajectories. \\

When treating the phonon mode quantum mechanically, single-phonon scattering terms manifest with coupling rates $\Gamma_- = g_0 \Omega_0 \sqrt{n+1} / \omega_\text{SAW}$ and $\Gamma_+ = g_0 \Omega_0 \sqrt{n} / \omega_\text{SAW}$, corresponding to removing and adding single phonons from the resonator, respectively. This originates from a perturbative expansion of the coupling Hamiltonian in the normalized rate $g/\omega_m$, and so applies formally only to small phonon occupancies. See reference 53 of the manuscript for further details. For large $n$, these rates are similar and we refer to them both simply as $\Gamma$ in the manuscript. \\

We numerically calculate the dynamics of the system, described by Eqn. 2, using the Lindblad master equation for the density matrix $\rho(t)$:
\begin{equation}
\dot{\rho}(t) = -\frac i\hbar [\tilde{\hat{H}}(t), \rho(t)] + \frac 12 \sum_n [2 \hat{C}_n \rho(t) \hat{C}_n^\dagger - \rho(t) \hat{C}_n^\dagger \hat{C}_n - \hat{C}_n^\dagger \hat{C}_n \rho(t)]
\end{equation}

\noindent with collapse operators $\hat{C}_n$=$\sqrt{\gamma_n} \hat{\sigma}_n$ and rates $\gamma_n$. Radiative decay is included through the Lindblad operator $\hat{\sigma}_- = (\hat{\sigma}_x+i\hat{\sigma}_y)/2$ with rate $\gamma_\text{QD}$. Additional pure dephasing is included through the Lindblad operator $\hat{\sigma}_z$ with rate $\gamma_z$. Note that ``slow" spectral fluctuations --- those occurring over timescales significantly longer than the QD's lifetime --- affect the measured TLS trajectories in a qualitatively different way than pure dephasing. We did not account for such spectral fluctuations in our calculations. Calculations are performed using the open-source Python library QuTiP. \\

\subsection{Calculation details}
In Figs. 1b,c of the manuscript we present the dynamics in two different pictures. In both panels, we assume an initial state $\ket{g,n}$, a total evolution time of 1.7 ns, and no dephasing. In Fig. 1b, we calculate the occupancy of the state $\ket{e,n}$ by calculating TLS dynamics between states $\ket{g,n}$ and $\ket{e,n}$ with Rabi rate $\Omega_0/2\pi$=1 GHz and $\Delta/2\pi$=$-\omega_\text{SAW}/2\pi$=$-$3.5 GHz (black curve). We then calculate the occupancy of the state $\ket{e,n-1}$ by calculating TLS dynamics between states $\ket{g,n}$ and $\ket{e,n-1}$ with the renormalized interaction rate $\Gamma$=$g_0 \Omega_0 \sqrt n / \omega_\text{SAW}$ with $\Delta$=0, $\Omega_0/2\pi$=1 GHz and $g_0\sqrt n$=$2\pi\times$1 GHz. These simplifications are largely for illustrative purposes, although they may be considered accurate in the artificial case of no dephasing. In Fig. 1c, we ignore the discrete phonon levels and calculate the QD's excited state occupancy by calculating the dynamics between the bare states $\ket g$ and $\ket e$ with $\Omega_0/2\pi$=1 GHz, $\Delta/2\pi$=$-\omega_\text{SAW}/2\pi$=$-$3.5 GHz and optomechanical coupling rates $g$=0 (black trajectory) and $g/2\pi$=2 GHz (red trajectory). Calculations in the manuscript Fig. 3 are like the ones shown in Fig. 1c. We assume an initial QD state $\ket g$, detuning $\Delta$=$-\omega_\text{SAW}$$\approx$$-$3.5881 GHz, Rabi rates $\Omega_0$ corresponding to measured pump amplitudes (specified in the manuscript) and $g$ values corresponding to measured microwave powers (described further in the calibration procedure below). 

\subsection{Mechanical phase}
\begin{figure}
      \includegraphics[width=0.8\textwidth]{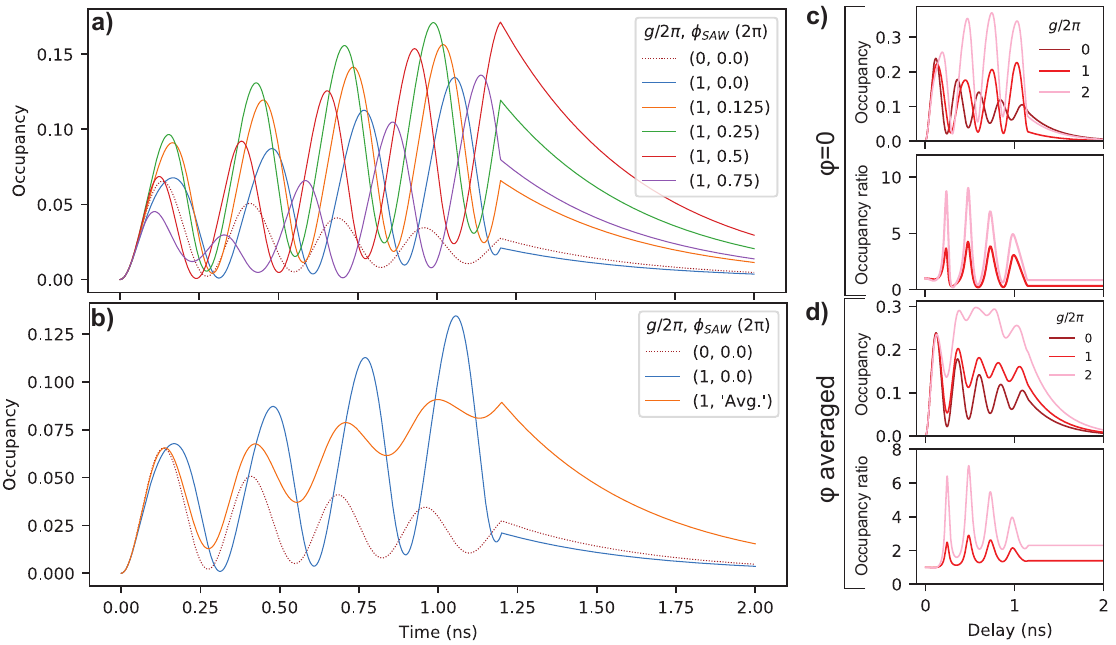}
	\caption{
	(\textbf{a}) Calculated TLS occupancies for five different values of $\phi$ (solid curves). The $g=0$ trajectory (which does not depend on $\phi$) is also shown for reference (dotted curve) 
	(\textbf{b}) Calculated TLS occupancies for $\phi$=0 (blue curve) and for $\phi$ averaged over eight uniformly-spaced values between 0 and $2\pi$ (orange curve). 
	All calculations in panels a,b were performed for an idealized square pulse with the following parameters: $g/2\pi$=1 GHz, $\Omega_0$=1 GHz.
	(\textbf{c,d}) Occupancies and occupancy ratios ($s_g(t) / s_{g=0}(t)$) calculated for the same set of parameters as in manuscript Fig. 3d for (c) $\phi$=0 and for (d) $\phi$ averaged over eight values between 0 and $2\pi$.
	}
\label{fig:SAW_phase}
\end{figure}

The mechanical phase $\phi$ affects the specific trajectory around the Bloch sphere. As described in the manuscript, our measurements sample many instances of $\phi$ and so the presented occupancy dynamics represent averages over all $\phi$. Calculations shown in manuscript Figs. 3c-d account for this by averaging calculated trajectories over eight uniformly spaced values of $\phi$ ranging from 0 to $2\pi$. Figs. \ref{fig:SAW_phase}a,b illustrate how the trajectory changes with $\phi$ (panel a), and also the result of averaging over these trajectories (panel b). Figs. \ref{fig:SAW_phase}c,d show that this averaging has only weak effects on the occupancy ratio $c_g(t)$ although the specific dynamics are changed.

\section{Sample Fabrication}
Our devices are fabricated on InAs quantum dot (QD) samples grown by molecular beam epitaxy (MBE) on a commercial GaAs wafer. The vertical structure consists essentially of a p-i-n diode above a distributed Bragg reflector (DBR) comprising 22 pairs of alternating AlAs/GaAs layers. Self-assembled InAs QDs are grown in the intrinsic (i) region between the highly doped p and n GaAs regions. The n (GaAs:Si at approximately $2\times10^{18}$ cm$^{-3}$) layer is 46.8 nm thick at a depth 300 nm below the surface. The InAs quantum dots (QDs) are grown 25 nm above this n layer. The p (GaAs:C at approximately $2.5\times10^{19}$ cm$^{-3}$) is 45 nm thick and lies at the sample surface. A 220 nm thick AlAs/GaAs digital alloy (current blocking layer) is grown between QDs and the p layer. 

The MBE-grown wafer is then processed in a cleanroom for QD p-i-n gate control and active surface acoustic wave (SAW) cavities. SAW mirror trenches are defined by electron beam lithography (EBL) and etched by reactive-ion etching (RIE). Large wirebond pads are defined for electrically contacting the n and p layers. Ohmic contacts to the n layer are made by etching vias around the SAW resonators down to approximately 80 nm above the n layer, and then depositing approximately 400 nm of AuGe/Ni/Au by electron-beam evaporation. Ohmic contacts to the p layer are made by depositing approximately 200 nm of Pt/Au to the p surface. The metals are thermally annealed in forming gas (95\% Ar and 5\% H2) at 430 C for 1 minute. The n regions around each SAW device are selectively removed by completely etching the surface to a depth 375 nm (25 nm below the n layer) so that this conductive layer does not interfere with on-chip microwave electronics. The n layer is retained only under SAW cavities (including mirrors and interdigital transducers, IDTs) and regions where the n contact metals were previously deposited. The coplanar waveguide (CPW) geometry is then defined by photolithography and formed by depositing 400 nm of Ti/Au followed by a lift-off process. The thickness of the CPW metal is chosen to be thicker than the etched depth so that it is connected across the n etch step. Finally, 20 nm of Al is deposited to make the IDTs. The IDT structure is defined by EBL and then formed using a lift off process. 

\section{Experimental Setup}
All of our QD optical measurements were performed using a 5 K optical cryostat with an internal cryogenic 0.81 NA objective. Nonresonant and resonant pump beams were shaped outside the cryostat using single-mode optical fiber components and free-space optical components. The resonant pump beam was a tunable external-cavity diode laser with an approximately 10 kHz linewidth at 960 nm. The nonresonant pump was a 632 nm diode laser. Pump beams illuminate the QD through the objective. QD luminescence was collected by the same objective and isolated from the pump predominantly by spectral filtering (for measurements involving a nonresonant pump) or polarization filtering (for measurements involving a resonant pump) with greater than 10$^7$ pump rejection. Luminescence was collected into single-mode optical fiber and various frequency- and time-domain measurements were performed using procedures detailed below. The QD bias voltage was held constant at 0.41 V for all measurements presented in the manuscript.

\subsection{Bias-dependent QD Photoluminescence Spectra}
\begin{figure}
      \includegraphics[width=0.4\textwidth]{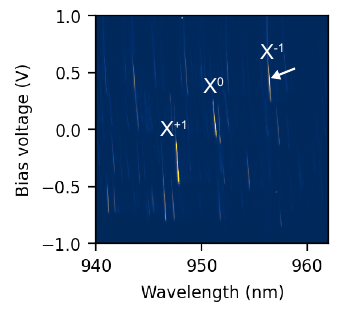}
	\caption{	
	QD PL spectrum with varying p-i-n bias voltage. Approximately three QDs contribute to this spectrum. The brightest three lines are associated with three distinct charge states of the single QD investigated in this work: $X^0$, neutral exciton; $X^{+1}$, single positively charged exciton; $X^{-1}$, single negatively charged exciton. The specific charge state investigated in this work is indicated by the white arrow.
	}
\label{fig:VBias_PL}
\end{figure}

QDs were illuminated with nonresonant pump light. Photoluminescence (PL) was collected into a single-mode fiber and reflected pump light was spectrally rejected. Pump powers were kept well below saturation levels. PL was sent to a spectrometer and recorded with a CCD camera. The QD bias was varied in discrete steps while repeated integrations were performed. Spectra were stitched together in post-processing. The bias-dependent PL spectrum from the QD interrogated in the manuscript is shown in Fig. \ref{fig:VBias_PL}.

\subsection{Continuous-wave Resonance Fluorescence Spectra}
\begin{figure}
      \includegraphics[width=0.4\textwidth]{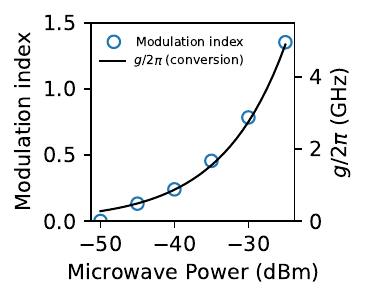}
	\caption{	
	Calibration between microwave driving power $P_\mu$ and coupling rate $g$. Orange markers: data. Blue curve: fit assuming $g \propto \sqrt{P_\mu}$.
	}
\label{fig:g_calibration}
\end{figure}

Continuous-wave (CW) resonance fluorescence (RF) spectra (e.g. manuscript Fig. 2d) were recorded by slowly sweeping the frequency of a tightly focused resonant pump beam over the QD’s exciton resonance while holding the QD bias constant. Luminescence counts at each frequency were recorded using a superconducting nanowire single photon detector (SNSPD) mounted in a dilution refrigerator. A pump power of approximately 500 pW at the sample was used, corresponding to approximately 1/10 the saturation level of the QD, providing a resonance fluorescence signal with approximately 150 kcounts/second when pumping the QD resonantly. For manuscript Fig. 2d (upper panel), such spectra were recorded without SAW driving and with SAW driving using an external coherent microwave source at approximately 3.5881 GHz and external microwave driving powers of approximately -41.5 dBm and -36.5 dBm. External microwave driving powers were converted to coupling rates $g$ by a series of calibration data where the modulation index $\chi$ was fit from optical spectra at each microwave driving power and $g$ was derived by the relationship $g$=$\omega_\text{SAW} \chi$. The calibration results are shown in Fig. \ref{fig:g_calibration}.

For manuscript Fig. 2d (lower panel), the ``resonant" pump was held at a constant frequency detuning of $\Delta$=$-\omega_\text{SAW}$. A pump power of approximately 2 nW was used. The frequency spectrum of scattered photons was measured by passing the collection port through a tunable Fabry-Perot etalon filter with a 600 MHz linewidth before being counted on an SNSPD. Such spectra were again measured without SAW driving and with SAW driving at 3.5881 GHz and the same microwave drive powers as in manuscript Fig. 2d (upper panel).

\subsection{Pulsed Optical Measurements and Time-domain QD Dynamics}
The time evolution of the QD dynamics was measured using standard time-correlated single-photon counting (TCSPC) techniques with a two-channel picosecond event timer and TCSPC module. For ``gradually decreasing" pulses, optical pulses were generated by passing a CW optical pump (same as described above) through two cascaded electro-optic modulators (EOMs). The pulse center frequency was detuned by $\Delta$ with respect to the QD’s resonance frequency (detuning specified in each measurement). The EOMs were driven by amplified pulses from an approximately 100-ps square-wave electronic pulse generator (EPG). The EPG was externally triggered using an arbitrary waveform generator (AWG). The rise-time, fall-time and width of the output pulses were measured to be approximately 15 ps, 15 ps, and 130 ps respectively. The on/off contrast was measured to be about $10^5$. The frequency bandwidth of the output pulses was measured to be approximately 7 GHz. To decrease the bandwidth (and increase the pulse duration), the pulses were transmitted through a 600 MHz tunable Fabry-Perot etalon which was tuned for maximum transmission at the pulse’s center frequency. The resulting temporal envelope is shown in Fig. 3a (top panel). For square pulses, the EOMs were directly driven using a high-speed AWG (64 GS/sec) with approximately 30 ps rise time and pulses of arbitrary length. The pulse center frequency was detuned by $\Delta$ with respect to the QD’s resonance frequency (detuning specified in each measurement). Measured pulse profiles shown in the manuscript may differ slightly from the pulse experienced by the QD due to small system reflections inside and outside the cryostat. The time distribution of the QD luminescence was measured by again detecting the luminescence with an SNSPD and time-correlating the counts on the TCSPC module with respect to a trigger signal from the AWG. In both pulse cases, the repetition rate was 34.085 MHz. We typically used 30-60 second total integration times, over which we collected typically tens of millions total photons. This type of measurement was performed with a wide range of average optical input powers (between approximately 0.25 nW to 300 nW). Pump powers were converted to resonant Rabi rates by looking at the resulting time-domain oscillations with zero detuning. 

The time-resolved QD dynamics shown in the manuscript are presented after subtracting the experimentally measured pump contribution. Nonetheless, in all cases, unrejected pump contributions were $\lesssim$1:100 compared to the measured signal and negligibly impact our results.

\section{Estimating QD Occupancies from Experimental Time-Domain Data}
\begin{figure}[h]
      \includegraphics[width=0.7\textwidth]{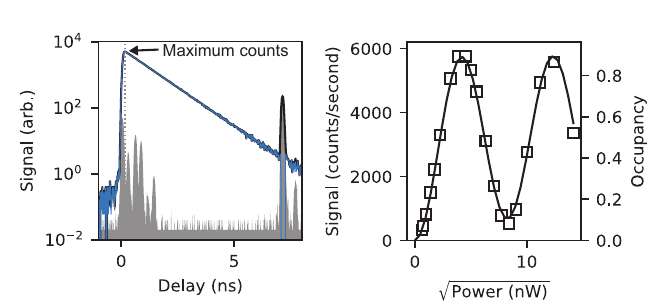}
	\caption{	
	Left panel: Raw TCSPC data (blue curve; units counts/second per time bin) acquired with $\Delta=0$ and a 130 ps pulse duration (gray region) and 16 ps time bin. ``Maximum counts" refers to the counts recorded in a 16-ps time bin at 140 ps, i.e., immediately after the pump pulse. Right panel: Counts recorded per second within a 16-ps time bin at 140 ps as a function of average optical power (open black square markers). Single-time-bin counts are converted to an expected QD occupancy $\langle \rho_{ee} \rangle$ by comparing the resulting Rabi oscillations to theoretical calculations including lifetime-limited dephasing at rate $\gamma_\text{QD}/2\pi$=320 MHz (solid black curve). 
	}
\label{fig:occupancy_calibration}
\end{figure}

QD occupancies were derived by comparing count rates from time-resolved QD luminescence data to a reference count rate. The reference count rate was derived by performing a set of time-resolved QD luminescence measurements with 130 ps optical pulses and widely varying optical pump powers and analyzing the resulting Rabi oscillations in the signal. Experimental Rabi oscillations were compared to numerical calculations of the QD occupancy under a 130-ps square pulse. The calculation assumes that the QD experiences no additional dephasing beyond the radiative lifetime limit. (The measured 1 ns radiative lifetime corresponds to a 321 MHz decay rate.) This calibration is illustrated in Fig. \ref{fig:occupancy_calibration}. Note that experimental occupancies reported here will be overestimated if the QD experiences additional dephasing. Nonetheless, excellent contrast observed in measured power-dependent Rabi oscillations (Fig. \ref{fig:occupancy_calibration}) indicate that additional dephasing beyond the radiative lifetime limit may be very small for this particular QD. This scaling is, however, irrelevant for the enhancements presented in the manuscript.

\begin{figure}[h]
      \includegraphics[width=0.95\textwidth]{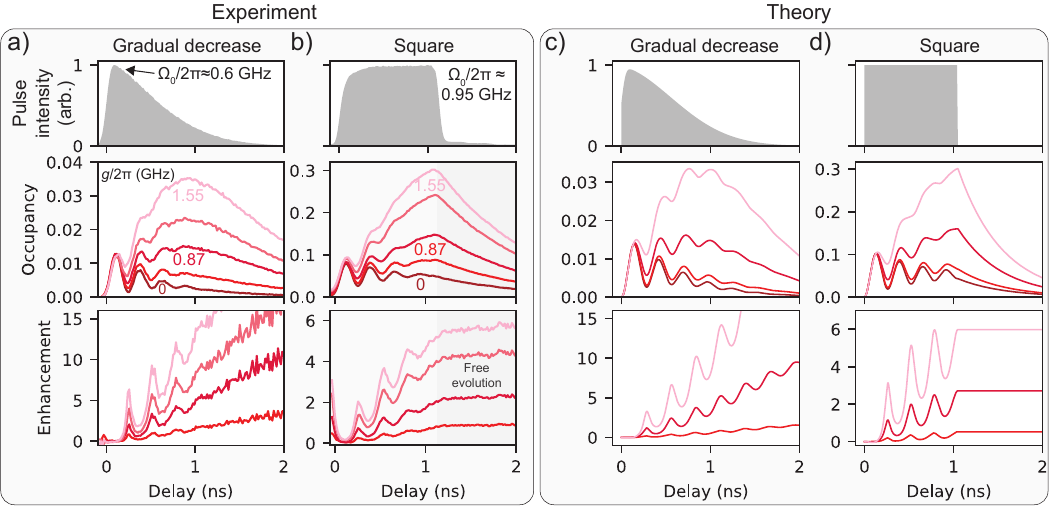}
	\caption{Data and calculations shown in Fig. 3 of the manuscript presented in 2D.
	}
\label{fig:oscillations}
\end{figure}

\begin{figure}[h]
      \includegraphics[width=0.5\textwidth]{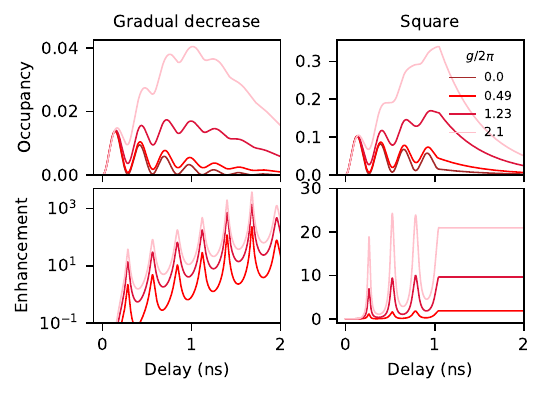}
	\caption{	
	Numerically calculated phonon-assisted QD excitation dynamics \emph{without} additional dephasing and no spectral drift. I.e., the QD dephasing is assumed to be radiative-lifetime limited with $\gamma/2\pi$=320 MHz. All other conditions are identical to those used for manuscript Figs. 3c,d. Left column: gradually decreasing pulse. Right column: square pulse. The ``enhancements" shown for the gradually decreasing pulse shape are presented in a log scale since maxima vary by several orders of magnitude over the pulse duration.
	}
\label{fig:oscillations_no_dephasing}
\end{figure}

\begin{figure}[h]
      \includegraphics[width=0.4\textwidth]{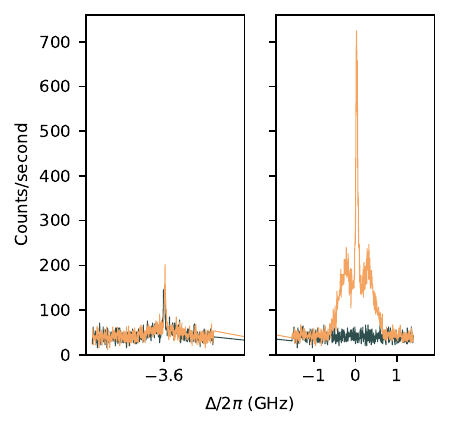}
	\caption{Scattering spectra from the same single QD excited with a continuous-wave narrowband red-detuned pump at $\Delta$=$-\omega_\text{SAW}$=$-$3.581 GHz. Spectra were measured with a frequency-tunable spectral filter with a 25 MHz transmission bandwidth. Gray curve: no SAW drive. Orange curve: SAW drive with $g$$\approx$1 GHz. The narrow ``coherent" Rayleigh scattering feature is seen at $\Delta$=$-$3.581 GHz in both spectra. When the SAW drive is on, some of this ``coherent" scattering component is frequency up-converted by phonons to leading to a narrow spike in the frequency at $\Delta$=0. This process also leads to a finite average QD occupancy and thus a broader luminescence feature with a width of approximately 1 GHz around $\Delta$=0. We suspect the small ``dips" between the narrow and broad emission features at $\Delta$=0 originate from phase interference between two scattering channels, but the origin is still under investigation.
	}
\label{fig:CW_SAWs}
\end{figure}

\end{document}